\documentclass[sn-mathphys-num]{sn-jnl}

\usepackage{graphicx}%
\usepackage{multirow}%
\usepackage{amsmath,amssymb,amsfonts}%
\usepackage{amsthm}%
\usepackage{mathrsfs}%
\usepackage[title]{appendix}%
\usepackage{xcolor}%
\usepackage{textcomp}%
\usepackage{manyfoot}%
\usepackage{booktabs}%
\usepackage{algorithm}%
\usepackage{algorithmicx}%
\usepackage{algpseudocode}%
\usepackage{listings}%
\usepackage{graphicx}
\usepackage{tabularx}
\usepackage{enumitem}
\usepackage{breakurl}
\usepackage{hyperref}
\usepackage{csquotes}
\usepackage{subcaption}
\usepackage{acronym}
\usepackage{nicefrac}
\usepackage{lmodern}
\usepackage{siunitx}

\newcolumntype{P}[1]{>{\centering\arraybackslash}p{#1}}

\newcommand{\gSet}[1]{\{\,#1\,\}}
\newcommand{\norm}[1]{\left\lVert\,#1\,\right\rVert}

\newcommand{\schedulevar}{U}

\newcommand{\targetvarcarrier}{c^C_{\rm target}}

\newcommand{\scheduleindex}{g}
\newcommand{\agentindex}{a}

\newcommand{\schedulelengthcarrier}{n^{\rm schedule}_{C,a}}

\newcommand{\agentlength}{n^{\rm agent}}

\begin{document}

\title[Coordination of Electrical and Heating Resources by Self-Interested Agents]{Coordination of Electrical and Heating Resources by Self-Interested Agents}


\author*[1,2]{\fnm{Rico} \sur{Schrage}}\email{rico.schrage@uni-oldenburg.de}
\author[1]{\fnm{Jari} \sur{Radler}}\email{jari.radler@uni-oldenburg.de}
\author[1,2]{\fnm{Astrid} \sur{Nieße}}\email{astrid.niesse@uni-oldenburg.de}

\affil*[1]{\orgdiv{Digitalized Energy Systems}, \orgname{Carl von Ossietzky Universität Oldenburg}, \orgaddress{\street{Ammerländer Heerstraße 114-118}, \city{Oldenburg}, \postcode{26129}, \state{Lower Saxony}, \country{Germany}}}
\affil[2]{\orgdiv{Energy Division}, \orgname{OFFIS -- Institute for Information Technology}, \orgaddress{\street{Escherweg 2}, \city{Oldenburg}, \postcode{26121}, \state{Lower Saxony}, \country{Germany}}}


\abstract{With the rise of distributed energy resources and sector coupling, distributed optimization can be a sensible approach to coordinate decentralized energy resources. Further, district heating, heat pumps, cogeneration, and sharing concepts like local energy communities introduce the potential to optimize heating and electricity output simultaneously. To solve this issue, we tackle the distributed multi-energy scheduling optimization problem, which describes the optimization of distributed energy generators over multiple time steps to reach a specific target schedule. This work describes a novel distributed hybrid algorithm as a solution approach. This approach is based on the heuristics of gossiping and local search and can simultaneously optimize the private objective of the participants and the collective objective, considering multiple energy sectors. We show that the algorithm finds globally near-optimal solutions while protecting the stakeholders' economic goals and the plants' technical properties. Two test cases representing pure electrical and gas-based technologies are evaluated.}




\maketitle

\section{Introduction}
Operating \ac{CHP} plants as \ac{DER}s in district heating grids while generating electrical energy is difficult for the operators due to the dependence of the energy forms on each other. Further, it can make sense to form \ac{VPP} \cite{bitsch2002virtuelle,niee2015fully} to gain access to the energy market while also needing to fulfill specific heating demands. These \ac{VPP}s can contain multiple \ac{CHP}s and \ac{RER}s operated by different actors. In most applications of energy scheduling within \ac{VPP}s, heating is often not part of the optimization. Besides \ac{VPP}s, there is also the upcoming topic of local energy communities \cite{koirala2016energetic} (and to an extent, it also holds for energy management of microgrids) in which the closed communities aim to be autonomous and self-sufficient. In these settings, optimal scheduling must deal with multiple stakeholders and heterogeneous power plants integrating electrical and thermal demand. Both applications require the agents to work together towards a common goal in a coalition.

In this work, we tackle the economic optimization problem of decentralized electrical and thermal units, which may be bound to each other due to \ac{CHP}s and heating pumps. To acknowledge the rise of decentralized power plants and prosumer-based district heating, it may be necessary to create schedules distributedly in one step (one optimization process) to avoid sharing private information. The second case is essential when assuming self-interested agents, which do not only focus on the coalition's objective to fulfill a target schedule, but also have a second private objective to optimize their utility. 

To consider all these difficulties and challenges, we present a fully distributed approach with a multi-level optimization, which considers private individual objectives while also focusing on the collective goal of all participating agents. We apply our optimization methodology to two scenarios, showing the approach's effectiveness and the difficulties of various types of participants. The first scenario utilizes \ac{CHP}s to fulfill the needs of the multi-agent system. In contrast, the second will rely on electric-only generators (\ac{HP}, \ac{WPP}, \ac{SPP}, and storage). In some scenarios, we depend on the recent flexibility model Amplify \cite{tiemann2022operational} to model storage flexibility and to find a feasible operational plan.

The work is structured as follows. After the introduction, we provide an overview of the related work. Then, the assumed system is described with the global objective and its modeling. This is followed by the power plants' models and the individual agents' economic objectives. After that, we present the distributed optimization heuristic and the novel multi-level decision process used to decide and generate the appropriate plant execution plan. Then, the evaluation methodology is described, and the results of the experiments are shown and discussed. Finally, we conclude with a summary, discussion, and an outlook.
\begin{figure*}[htb]
    \centering
    \includegraphics[width=0.6\textwidth]{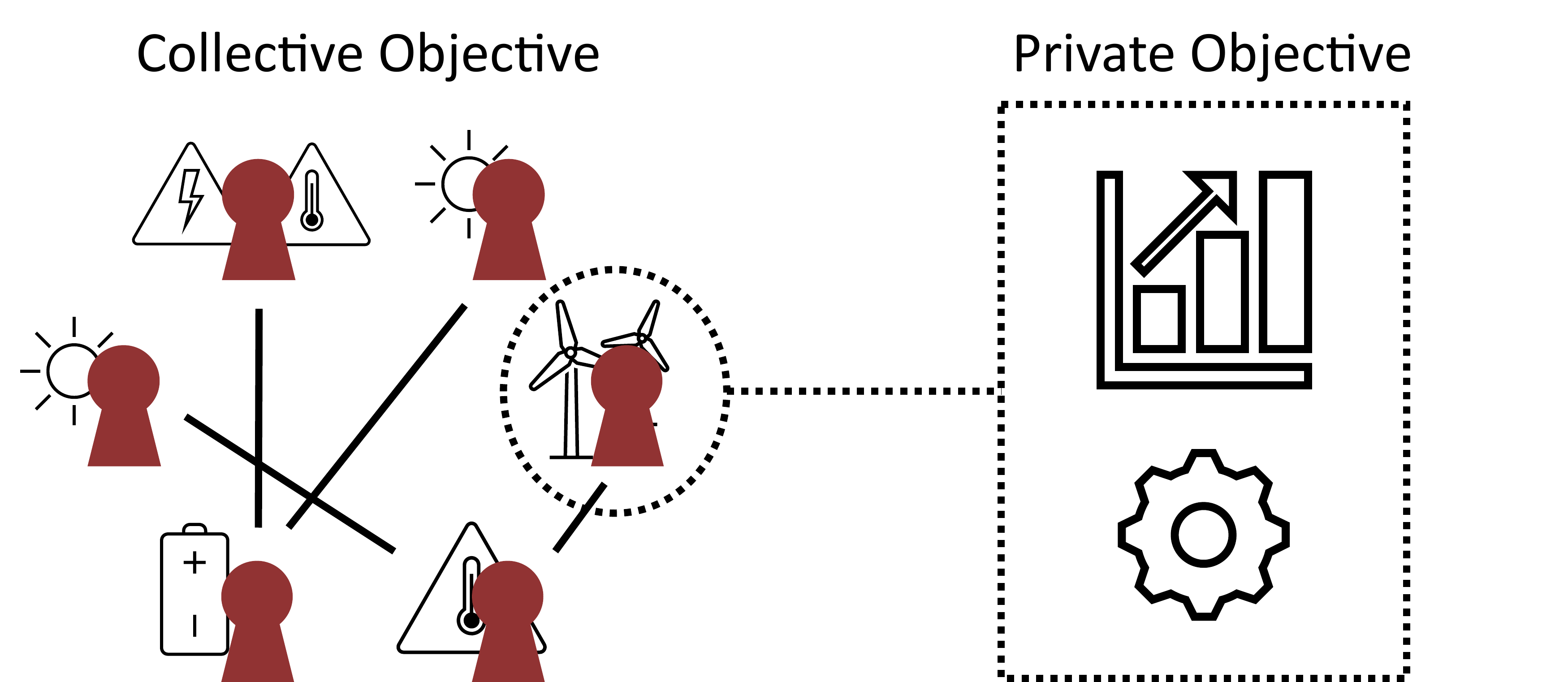}
    \caption{Overall system structure}\label{fig:system_model}
\end{figure*}
\section{Related Work}
This work focuses on distributed decisions for the distributed energy scheduling problem. We include private objectives for every individual actor (represented by an agent) and one global objective for the whole coalition. These objectives should not need to be convex for feasible solutions. Further, the agents shall not reveal the technical details of their physical components. There is also a natural time limit on the execution time, e.g., due to some applications like redispatch service provision or the short cycles of some markets.

Distributed energy scheduling is a highly relevant topic in the literature regarding electricity systems \cite{li2010coordination,en13153921,logenthiran2011multi,silva2012integrated}. In contrast, combined scheduling in electricity-heating systems has not yet been sufficiently tackled. However, electricity-heating scheduling can be considered a particular case of multi-objective optimization. Consequently, there are three relevant areas: literature on distributed multi-objective scheduling, distributed electrical energy scheduling, and distributed multi-energy scheduling. Note that there are central approaches with similar methods and similar problem formulations (e.g., \cite{DISOMMA2018272}), which are not addressed here.
\paragraph{Distributed Multi-Objective Scheduling} In \cite{schrage2023multi}, the authors developed an approach to integrate \ac{ES} in a distributed energy scheduling algorithm. For this, multiple objectives can be considered. However, the objectives are optimized locally and not in the distributed optimization approach, and they did not consider multiple energy carriers. However, the methodology is generally transferable as one objective can represent every carrier. Furthermore, in \cite{bremer2019evo}, the heuristic \ac{COHDA} \cite{hinrichs2017distributed} is extended to a multi-objective distributed algorithm using the evolutionary metaheuristic. This algorithm is extended and applied in the energy context by Stark et al. \cite{stark2024distributed}. They showed their approach can generate a solution of similar quality compared to the central NSGA-II \cite{deb2002fast} algorithm. Further, Pareto fronts are calculated, which slows down the calculation and often leads to suboptimal solutions for specific weights (given that the interesting weights are known). 
\paragraph{Distributed Electricity Scheduling} There are plenty of approaches in this area (\cite{logenthiran2011multi,logenthiran2010multi,silva2012integrated,en13153921}), but none can cope with multiple energy carriers. Besides, using well-known distributed mechanisms like \ac{DGD} \cite{ram2009distributed} or the \ac{ADMM} \cite{wei2012distributed} may generally not be possible due to constraining mathematical requirements for the objective function (most require at least convexity). In these terms, combinatorial methods \cite{landaburu2006optimal,li2010coordination,hinrichs2017distributed} can provide better flexibility and privacy advantages.
\paragraph{Distributed Multi-Energy Scheduling} There are some approaches considering metaheuristic optimization in IES (integrated energy system), which optimize the schedules of generators of heating and electricity combined (e.g., \cite{chenMultiregionCombinedHeat2024}. Further, in \cite{app10041214} and \cite{robustSynSchedulingChen}, the authors used \ac{ADMM} for distributedly solving the multi-energy scheduling problem while also considering network constraints using the modeling approach of the \ac{EH}. The problem formulation of this work is based on multiple three \ac{EH}, which already aggregate numerous components, which may not always be a feasible model for the individual actor. 

To sum up, based on the outlined research gap, the following provided as contributions, currently not visible in other research:
\begin{enumerate}
    \item Efficient cross-sectional energy scheduling on the distribution level
    \item Development of a distributed hybrid heuristic approach for multi-level optimization
    \item Presenting and comparing the application of the optimization to a gas-based and a renewable scenario
\end{enumerate}
\section{System description}\label{sec:system_desc}
The system of interest consists of agents responsible for each generation unit (without loss of generality). A visualization is depicted in Figure \ref{fig:system_model}. There are solar, wind, \ac{CHP}, and storage units. To act together, these agents form a coalition to coordinate themselves and maximize the overall profit (the specific application may vary). Therefore, each agent and its unit have specific technical details and economic objectives, respectively, technical and economic constraints. These technical details and constraints shall stay hidden. Above these specific objectives, the coalition's objective has to be fulfilled. This is essential for fulfilling the coalition/\ac{VPP} orders, so the agents must prioritize this global objective. 

In the first step, we describe the systems' global objective, which shall be maximized. The following equations define the agent systems' objective to fulfill multiple carriers' target schedules. 
\begin{equation}\label{eq:optimization_objective}
  \begin{split}
      U(C) &= -\norm{\targetvarcarrier - \sum_{\agentindex=1}^{\agentlength}\sum_{\scheduleindex=1}^{\schedulelengthcarrier} (\schedulevar^C_{\rm \agentindex,\scheduleindex}\cdot x^C_{\rm \agentindex,\scheduleindex})}_1\\
      \text{with } &\sum_{\scheduleindex=1}^{\schedulelengthcarrier}x^C_{{\rm \agentindex,\scheduleindex}} = 1\\
      &x_{{\rm \agentindex,\scheduleindex}}\in\gSet{0,\,1},~\agentindex=1,\,\dots,\,\agentlength,~\scheduleindex=1,\,\dots,\,\schedulelengthcarrier.
  \end{split}
\end{equation}
Here, $U(C)$ is the global utility regarding energy carrier C, $\targetvarcarrier$ is the target schedule for energy carrier C, $\agentindex$ identifies a specific agent, $\agentlength$ is the total number of agents, $\scheduleindex$ identifies a specific possible schedule of an agent, $\schedulelengthcarrier$ is the total number of schedules for a particular agent and carrier, $\schedulevar$ is the $\scheduleindex$'th schedule of agent $\agentindex$ regarding carrier C, $x^C_{\rm \agentindex,\scheduleindex}$ is the selection of the $\scheduleindex$'th schedule of agent $\agentindex$ regarding carrier C.

We formulate a maximization problem based on this formulation of the scheduling utility.
\begin{equation}\label{eq:generic_optimization_objective}
      \underset{\zeta\in \mathcal{C}\colon x_{a,\,x}^\zeta}{\text{max}}~\sum_{\zeta\in\mathcal{C}}U(\zeta)
\end{equation}
Here, $\zeta$ is the carrier, and $\mathcal{C}$ is the set of carriers included in the abstract optimization problem.

In this work, heat and electricity are part of the scheduling problem formed by $U(P) + U(H)$. $P$ represents electricity carrier and $H$ the heat carrier.
\begin{equation}\label{eq:this_optimization_objective}
\underset{x^P_{\rm\agentindex,\scheduleindex},\,x^H_{\rm\agentindex,\scheduleindex}}{\text{max}}~\left(U(P) + U(H)\right)
\end{equation}
\section{Modeling of components and objective}
In the optimization, we will use different types of generators to cover a huge proportion of possibilities, as one of the core contributions is enabling the collaboration between different types of generators of different carrier types. We focus on the following types.
\begin{itemize}
    \item Combined heat and power
    \item Renewables, \ac{WPP} and \ac{SPP}
    \item \ac{HP}
    \item \ac{TES} and electrical storage
\end{itemize}
\subsection{CHP}
The \ac{CHP} is modeled using efficiency and energy weights, which result in the conversion of a specified amount of gas to power and heat. Further, we define the maximum amount of gas consumed for every timestep and the maximum amount of electrical power curtailed (or used otherwise) per timestep. 
\begin{equation}
    \begin{split}
        P_{\rm CHP} &= \rho {\frac{\dot{M}\cdot\rm HHV}{3600}} \\
        H_{\rm CHP} &= \eta {\frac{\dot{M}\cdot\rm HHV}{3600}} \\
        \text{with } &\forall i: 0 \leq \dot{m}_i \leq \dot{m}_{\rm max} \text{, } 0\leq i < 96 \\
        \text{and } P_{\rm CHP} &- P^{\rm actual}_{\rm CHP} \leq P^{\rm reserved\,max}_{\rm CHP}
    \end{split}
\end{equation}
Here, $P_{\rm CHP}$ is the power output of the \ac{CHP} in \unit{W}, $H_{\rm CHP}$ is the heating power output of the \ac{CHP} in \unit{W}, $\rho$ is the power generation efficiency, $\eta$ is the heating generation efficiency, $\dot{M}$ is the vector of gas mass consume rates in \unit{kg/h}, HHV is the higher heating value of the gas being used (in \unit{J/kg}), $\dot{m}_i$ is the consumed gas flow rate at step $i$ in \unit{kg/h}, $P^{\rm actual}_{\rm CHP}$ is the power output in \unit{W} used for the optimization, $P^{\rm reserved\,max}_{\rm CHP}$ is the maximum power in \unit{W}, which can be used for other purposes.
\begin{figure*}[tb]
    \centering
    \begin{subfigure}{0.62\textwidth}
        \centering
        \includegraphics[width=1\textwidth]{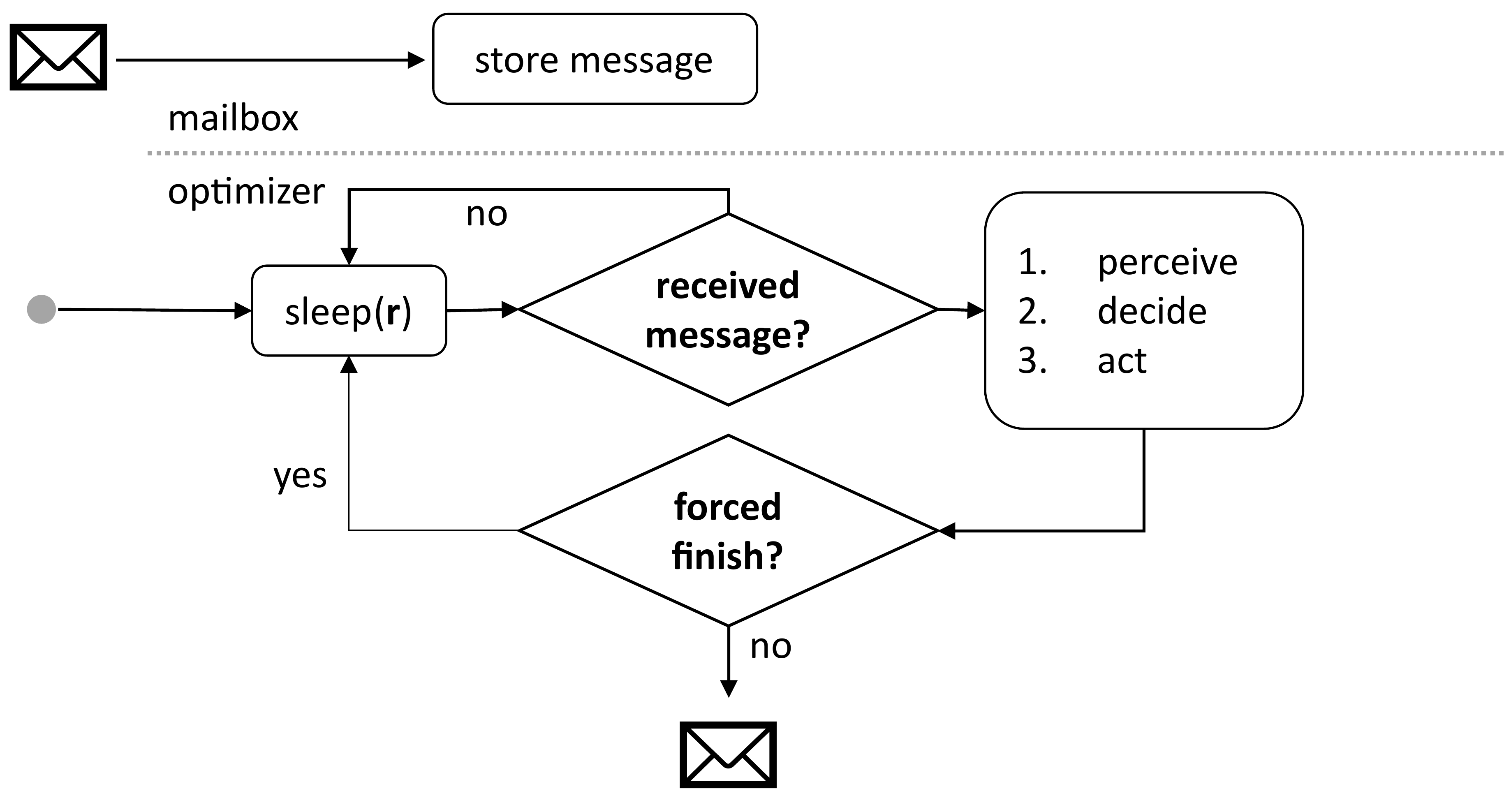}
        \subcaption{Overall execution flow of an agent}\label{fig:overall_flow}
    \end{subfigure}
    \begin{subfigure}{0.37\textwidth}
        \centering
        \includegraphics[width=1\textwidth]{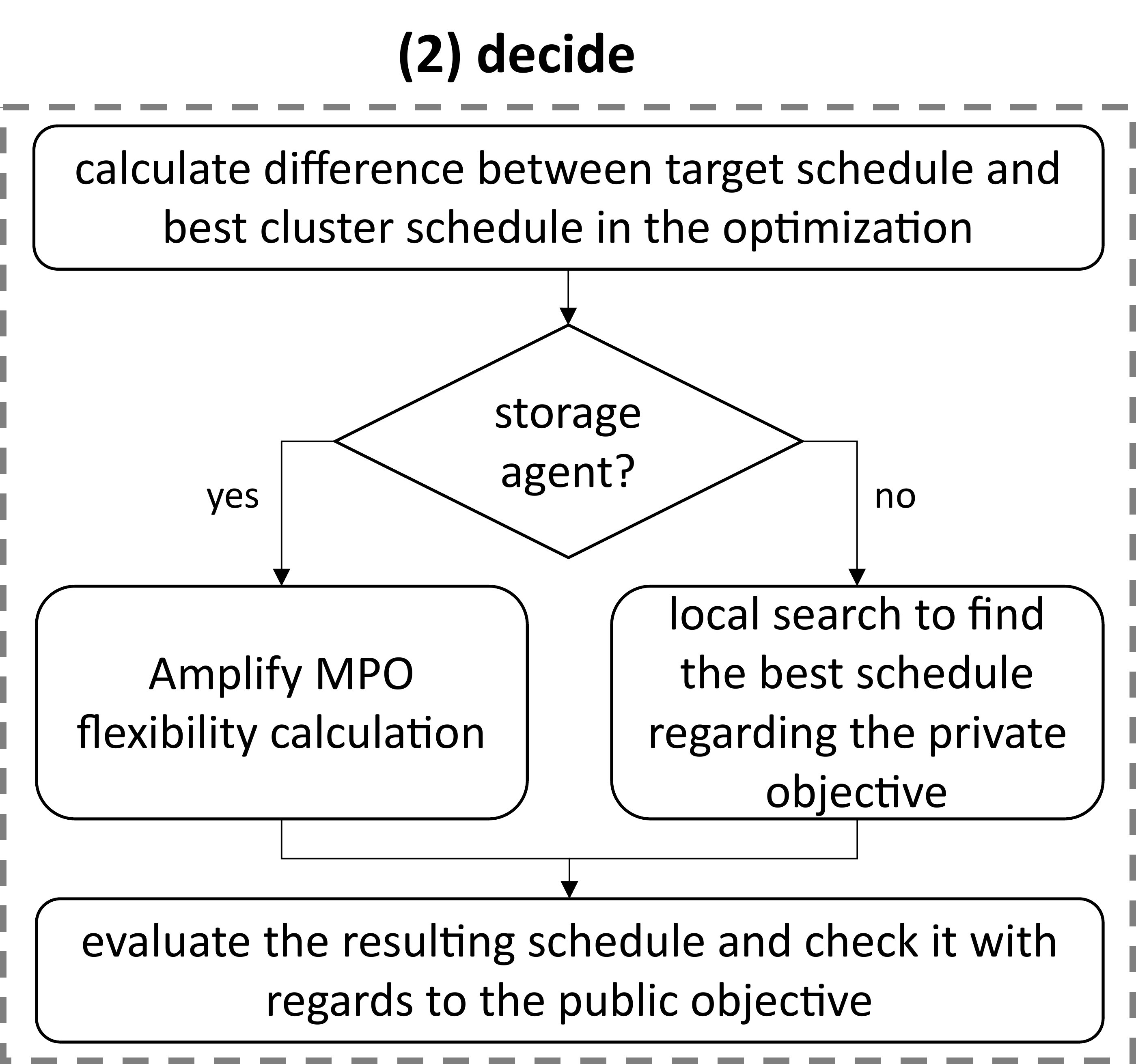}
        \subcaption{Decision process of an agent}\label{fig:decide_flow}
    \end{subfigure}
    \caption{Control flow of an individual agent}
\end{figure*}
\subsection{Renewables}
We model renewables as generated time-series data based on actual data samples for \ac{WPP}s and SPPs, which are explained as follows. 

The \ac{WPP} power per timestamp is calculated using the maximal wind speed over a day. The specific power in step $i$ is estimated by drawing a random sample from a normal distribution with the center $\sigma=3.7$ and the variance $0.5$; this parameter leads to the distribution of the wind speed change every 30min for the average wind speed in from an example German city (Rotenburg \cite{weatherRotenburg} in March 2023). We can calculate the power using the calculation equation for wind turbines by \cite{scheurich2016improving} and adding the drawn sample $\delta_w$ as follows.
\begin{equation}
    \nicefrac{1}{2}\rho AV^3K_P + 2\delta_w
\end{equation}
Here, $A$ is the rotor surface area, $\rho$ is the air density, and $V$ is the wind speed and the power coefficient $K_P$.

The \ac{SPP}s are generated using a solar profile from the same area (Rotenburg). The schedule is altered for every plant by adding random noise.
\subsection{Heat pump}
The \ac{HP} is modeled using a constant efficiency factor applied to convert from electrical energy to thermal energy. Here, the factor is chosen as an average of the day, as \ac{HP} efficiency depends on the external temperature.
\subsection{Storage}
The storage model of Amplify \cite{tiemann2022operational} is used. This model consists of the storage capacity, the maximum charge/discharge power, and the charge/discharge efficiencies. Self-discharge is neglected. The capacities of the used storage are chosen depending on the target flexibility of the scenario. Time can be considered discrete.

Amplify, as a flexibility model, is used in this work to calculate a schedule for energy storage given the necessary power capacities for every timestep. This is done by representing the storage with an upper and a lower flexibility band, limiting the available power of the storage for every step. For this, the following represents these bands \cite{tiemann2022operational}:
\begin{align}
    P^{\rm max}_{\rm avail}(i) &= min\left(P^{\rm max}_{\rm ch}(i),\,P^{\rm mpo}_{\rm dis}(i)\right)\\
    P^{\rm min}_{\rm avail}(i) &= max\left(P^{\rm max}_{\rm dis}(i),\,P^{\rm mpo}_{\rm ch}(i)\right)\\
    &\text{with } i\in\mathbb{N}, 0\leq i\leq i^{\rm max}
\end{align}
Here, $i$ is the interval, $i^{\rm max}$ is the number of intervals, $P^{\rm max}_{\rm ch}$ is the maximum charge power, $P^{\rm mpo}_{\rm dis}$ is the discharge power reserved for a so-called multi-purpose obligation, $P^{\rm max}_{\rm dis}$ is the maximum discharge power, and $P^{\rm mpo}_{\rm ch}$ is the reserved charge multi-purpose obligation. Note that these equations represent the band without considering the SoC; Amplify will constrain this band further based on the available energy, limiting the requested obligation further. For the exact information on how this is done, we refer to the original publication \cite{tiemann2022operational}.

In our approach, one multi-purpose obligation for every interval $i$ will be set to the difference between the currently scheduled power (the sum of all chosen schedules) and the power target. 
\subsection{Economic objective}\label{sec:individual_obj}
We model self-interested agents, which participate in the distributed optimization. Hence, every agent has an objective defined by their economic interests, capabilities, and private operation cost information. To consider all variables, we formulate a generalized utility function, which maps the current selection within the cluster schedule to costs. 
\begin{equation}\label{eq_private_objective}
    \begin{split}
        u(a) &= \gamma_a \cdot P_a + \delta_a \cdot H_a - p_{\rm all}(a) \\
        p_{\rm all}(a) &= p_P \cdot {\left|P_a - \Delta P\right|}^{p_e} - p_H \cdot {\left|H_a - \Delta H\right|}^{p_e} - \theta_a \cdot F_a  
    \end{split}
\end{equation}
Here, $u(a)$ is the objective function for agent $a$, $\gamma_a$ is the utility coefficient for the contributed electrical power of agent $a$, $P_a$ is the contributed power, $\delta_a$ is the utility coefficient for the contributed thermal power, $H_a$ is the contributed thermal power. As the agent's primary goal is the success of the distributed optimization they are participating in, the agent needs to consider the global objective $U(C)$ as well. This is done using this objective as a penalty in the individual objective function, as shown in equation \ref{eq_private_objective}. In this equation, $p_{\rm all}$ is the penalty,  $p_P$ is the penalty coefficient for the electrical power, $\Delta P$ is the difference between the electrical cluster schedule (sum of all agents schedules) and the electrical target schedule, $p_e$ is the penalty exponent, $p_H$ is the penalty coefficient for the thermal power, $\Delta H$ is the difference between the thermal cluster schedule and the thermal target schedule, $\theta_a$ is the resource cost coefficient and $F_a$ is the vector of contributed resources.

The coefficients can be vector or scalar, depending on the time resolution. The penalty should be selected to dominate the economic value for significant deviations from the target flexibility.
\section{Optimization approach}
After we have defined the individual (subsection \ref{sec:individual_obj}) and group objectives (section \ref{sec:system_desc}) and constraints, we need to specify the distributed optimization routine to optimize electrical and thermal generators/loads simultaneously. In this section, explain the gossiping heuristic on a fundamental level and the multi-level decision routine of the individual agents for the storage and non-storage units.

We base our approach on an adjusted version of the gossiping-based COHDA \cite{hinrichs2017distributed} heuristic for our distributed algorithm. The general flow of the adjusted heuristic is shown in Figure \ref{fig:overall_flow}. Every agent has a mailbox, which will be checked every $r$ milliseconds. The agents are connected using a static topology. Here, we assume a fully connected topology. If there is a new message from a connected agent, the three steps, (1) perceive, (2) decide, and (3) act, are executed. The messages contain the decision of the sending agent ($U^C_{a,g}$ for every $C$) and its knowledge about the decision of the other agents. This message is processed and integrated into the understanding of the receiving agent. We refer the interested reader for more details of this step to the original paper \cite{hinrichs2017distributed}. After the message is processed, the agent can decide (2) whether a new decision has to be made. If the agent calculated a new decision, this decision will be sent to all connected agents (3).

In the original publication, the decision step (2) searches for the best schedule in the fixed schedule set by the agent. This is not feasible for our use case, as we include storage units, which can not be modeled this way without tremendous performance restriction (also experimentally shown in \cite{schrage2023multi}). Further, our problem definition demands a simultaneous optimization of multiple interdependent sectors. We must somehow represent the flexibilities; while \cite{hinrichs2017distributed} used schedule sets, there are more complex ways, like representing the power plant using support vectors \cite{5953329}. We chose Amplify for storage as it includes the storage-specific parameters and handles the state (SoC) natively. For the other power plants, the flexibility representation is trivial due to the level of detail of these models. However, choosing an exact schedule is still a demanding task that we tackle with an optimization approach.

Consequently, we present our newly developed decision-making (step (2)) of the agent, which primarily determines the type of plants that can realistically participate and how well the resulting schedule performs. The overall flow of the decision algorithm is depicted in Figure \ref{fig:decide_flow}.

The input for the decision routine is the current collective solution $S^C_c$ (cluster schedule), the sum of all agents' single solutions $S^C_a$ (agent schedules). A solution is an ordered list of power values with a length of 96, one value per 15 minutes, resulting in a schedule with the planning horizon of a day.

First, the input to the routine is used to calculate the difference between the target schedules $c^C_{\rm target}$ and the collective solution $S^C_c$.
\begin{equation}
    \Delta S^C_c = c^C_{\rm target} - S^C_c        
\end{equation}
Based on these deltas, one of two different paths shall be chosen. 

If the agent controls storage, the Amplify \cite{tiemann2022operational} flexibility calculation is triggered. The multi-purpose obligations (MPOs) Amplify uses for the calculation can be set to $S^C_c$, while $C$ is equal to the storage carrier. The details of this calculation can be found in \cite{tiemann2022operational}. The feasible power band is directly used to find the minimal or maximal power feasible output of the storage for every timestep. The value of this newly generated schedule is set to $p_{\rm all}(a)$ with $a$ as the storage agent.

For all non-storage components, a local search optimization is executed. The local search is performed for every timestep $i$; the search is initialized with a random value $v_r$ between $0$ and min($P_{max}$, $\Delta S^C_{c, i} - S^C_{a, i}$). Then, the private utility is calculated for this value, and the values $v_r+1/v_r-1$; the utilities of these three options $O_1$, $O_2$, $O_3$ are computed. Further, the values are sorted by their utility descending. Then, we differentiate between four cases.
\begin{enumerate}
    \item If the utilities are strictly monotone increasing, change the lower bound to $O_2$ 
    \item If the utilities are strictly monotone decreasing, change the upper bound to $O_2$ 
    \item If $O_2 > O_1$ and $O_1 > O_3$ or $O_3 > O_1$ and $O_1 > O_2$, change the lower bound to $max(O_2,\,O_3)$ and the upper bound to $min(O_2,\,O_3)$
    \item The bounds are not modified
\end{enumerate}
\begin{table}[htb]
    \centering
    \begin{tabular}{cccccccccc}
    \toprule
    type & count & power (kW) & efficiency & capacity (kWh) \\\midrule
    solar & 15 & 0.2 & n.a. & n.a. \\
    \ac{CHP} & 6 & 1.1/0.7 & 0.45/0.40 & n.a. \\
    HS & 2 & 1.5 & 0.99 & 12/16 \\
    ES & 2 & 1.5 & 0.95 & 12/16  \\\bottomrule
    \end{tabular}
    \caption{Composition of the gas-based scenario}
    \label{tab:gas_scenario}
\end{table}
\begin{table}[htb]
    \centering
    \begin{tabular}{cccccccccc}
    \toprule
    type & count & power (kW) & efficiency & capacity (kWh) \\\midrule
    solar & 15 & 1.1 & n.a. & n.a. \\
    wind & 5 & $\sim 2$ & n.a. & n.a. \\
    \ac{HP} & 3 & 2.7/2.0/1.3 & 4 & n.a. \\
    HS & 1 & 10 & 0.99 & 84 \\
    ES & 1 & 10 & 0.95 & 84  \\\bottomrule
    \end{tabular}
    \caption{Composition of the pure electrical scenario}
    \label{tab:electric_scenario}
\end{table}
After the new bounds have been set, a new random value between the bounds is picked, and the procedure repeats until $n_{\rm iterations}$ or if the value would be out of bound otherwise. 

The result of this search will be a schedule with the locally optimal values at every step. Afterward, the global utility is calculated of this schedule and compared with the previous utility (or $-\infty$ if the first schedule is generated and selected). The schedule can be chosen if the global utility has improved. 
\section{Case Study}
To evaluate the approach, we developed two settings: (1) a gas-based setting using \ac{CHP}s and (2) a pure-electrical setting using \ac{HP}s. We chose these two settings to represent two common approaches to generating thermal energy. Each setting has different properties. While the \ac{CHP} generates both types of energy simultaneously, the heat pump consumes electrical energy and generates thermal energy. These results in structurally different necessary schedules, which is a neat property to evaluate our approach. Further, we vary parts of the objective, which results in different variants, shown in table \ref{tab:scenarios_overview}.

\begin{table}
    \centering
    \begin{tabular}{cp{8cm}}
    \toprule
         name & description \\
         \midrule
         GB & Gas-based scenario without any energy storage \\
         GBS-H & Gas-based scenario with energy storage and high penalties \\
         GBS-L & Gas-based scenario with energy storage and low penalties \\
         GBS-M & Gas-based scenario with energy storage including opportunity costs for the agent \\
         PES-H & Pure-electrical scenario with energy storage and high penalties \\
         PES-L & Pure-electrical scenario with energy storage and low penalties \\
         PES-M & Pure-electrical scenario with energy storage including opportunity costs for the agent \\\bottomrule
    \end{tabular}
    \caption{All scenarios variants listed and named}
    \label{tab:scenarios_overview}
\end{table}

The opportunity cost and penalty variants are used to determine the influence of changes in the utility function regarding the profit calculation and the weighting of the global utility in the private utility calculation.

The gas-based scenarios consist of 25 agents in total. \ac{SPP}s, combined heat power plants, heat storage, and electrical storage (expect for GB) participate in the scenario. The exact composition and their parameterization are shown in Table \ref{tab:gas_scenario}. The scenario represents gas-based optimization scenarios with low storage capacities and a low share of renewables.

The pure-electrical scenarios also consists of 25 agents. Instead of \ac{CHP}s, there are \ac{HP}s and multiple \ac{WPP}s for the time without solar energy. The most significant differences to the first scenario are the large electrical and thermal storage capacity and the 100\% share of renewable power plants. The exact composition and its parameters are shown in Table \ref{tab:electric_scenario}.

We will consider the robustness and effectiveness of the approach. The median, average, standard deviation and outliers are considered for robustness. The effectiveness will be measured with the fulfillment rate in percent of the electrical and heating target schedules ($c^{\rm el}_{\rm target}$ and $c^{\rm heat}_{\rm target}$).
\subsection{Implementation}
The simulation uses mango and the mango-library \cite{SCHRAGE2024101791} to implement the different agents for creating a multi-agent system and starting optimization negotiations. The storage is modeled, and its flexibilities are calculated using Amplify \cite{tiemann2022operational}.
\begin{figure}[h!tb]
    \centering
    \begin{subfigure}{0.49\textwidth}
      \centering
      \includegraphics[width=1\textwidth]{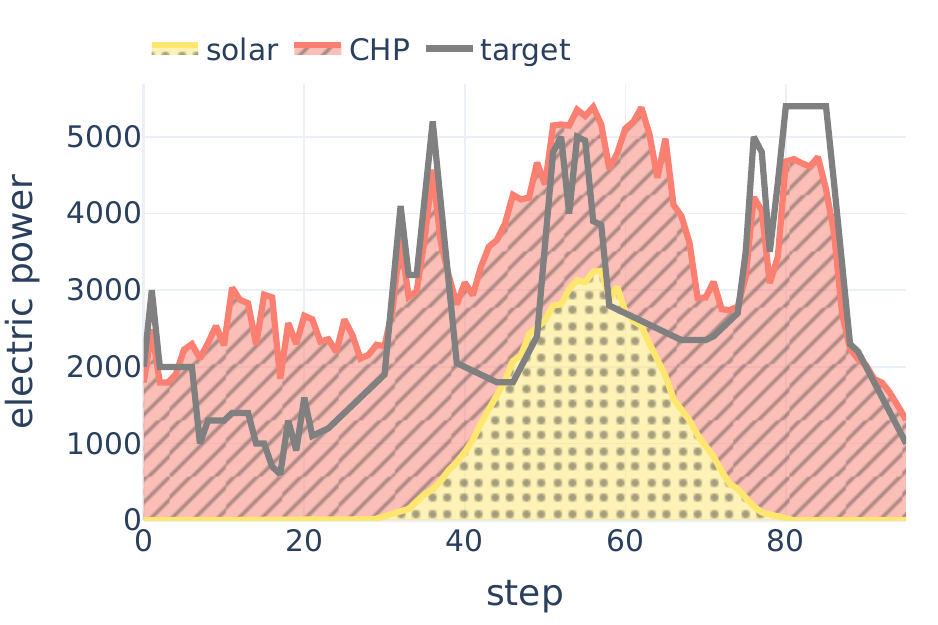}
      \subcaption{GB: Area plot of the power cluster schedule stacked by agent type}\label{fig:power_cs_industry}
    \end{subfigure}
    \begin{subfigure}{0.49\textwidth}
      \centering
      \includegraphics[width=1\textwidth]{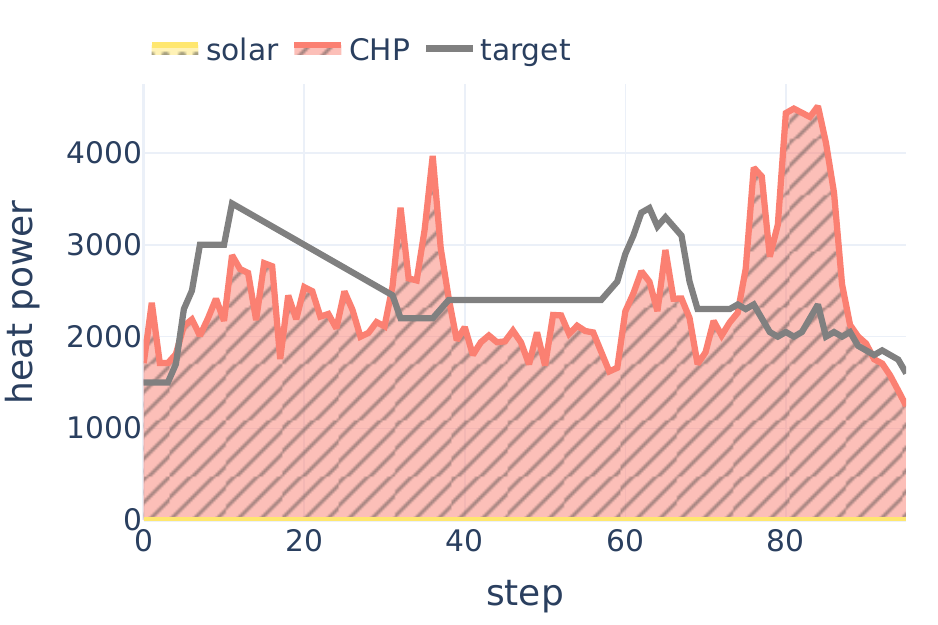}
      \subcaption{GB: Area plot of the heat cluster schedule stacked by agent type}\label{fig:heat_cs_industry}
    \end{subfigure}
    \begin{subfigure}{0.49\textwidth}
      \centering
      \includegraphics[width=1\textwidth]{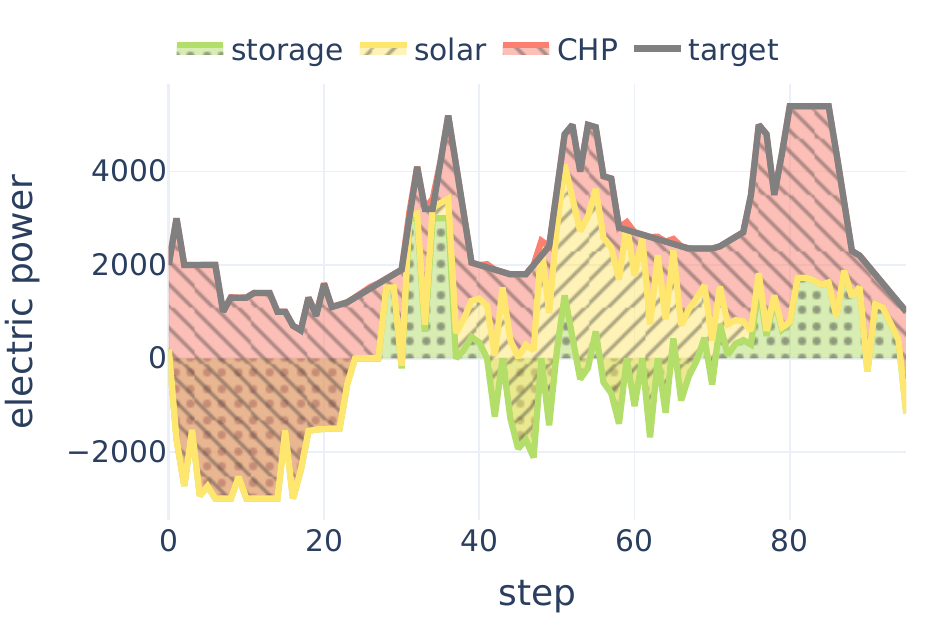}
      \subcaption{GBS-H: Area plot of the power cluster schedule stacked by agent type}\label{fig:power_cs_storage}
    \end{subfigure}
    \begin{subfigure}{0.49\textwidth}
      \centering
      \includegraphics[width=1\textwidth]{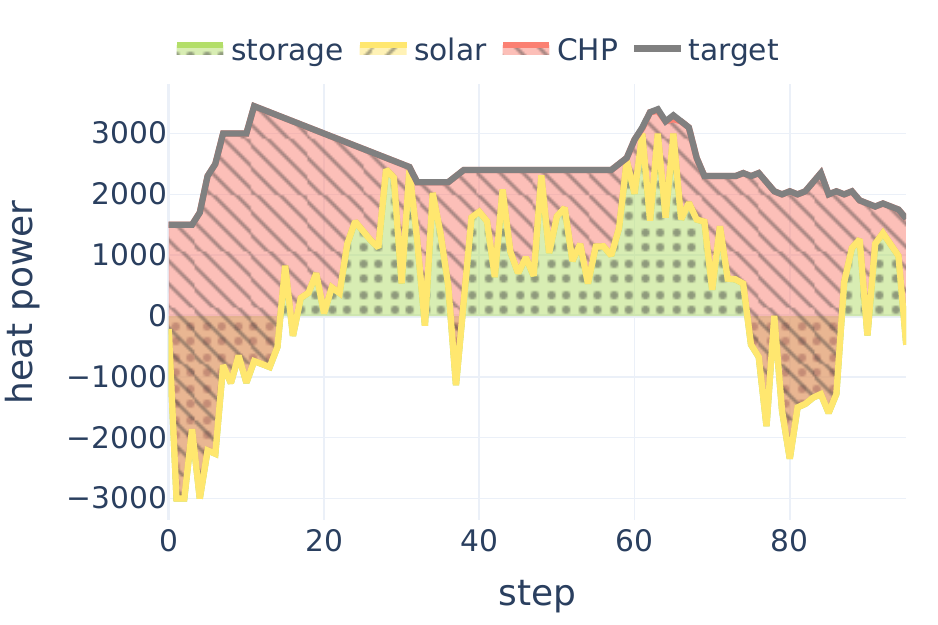}
      \subcaption{GBS-H: Area plot of the heat cluster schedule stacked by agent type}\label{fig:heat_cs_storage}
    \end{subfigure}
    \begin{subfigure}{0.49\textwidth}
      \centering
      \includegraphics[width=1\textwidth]{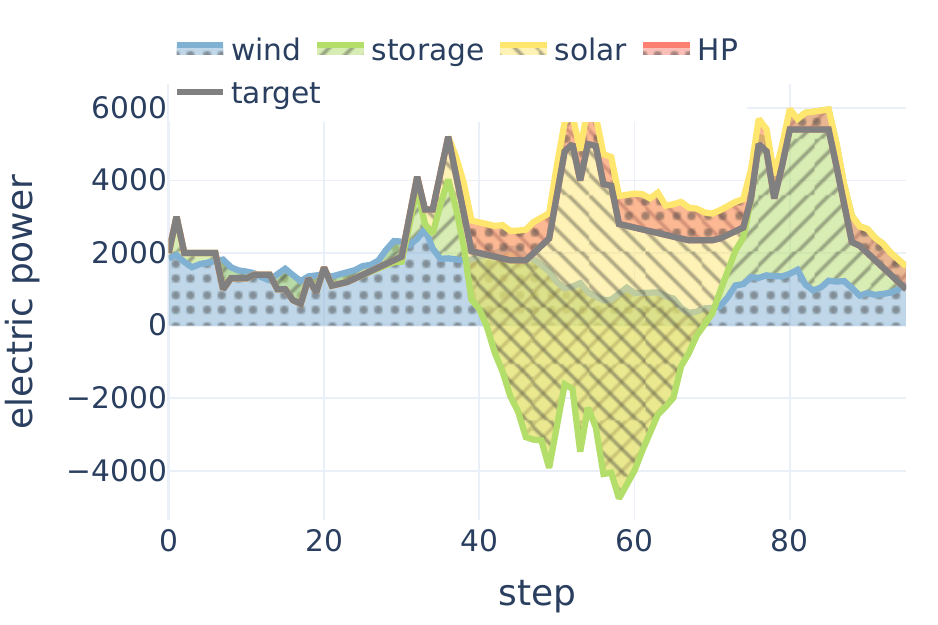}
      \subcaption{PES-H: Area plot of the power cluster schedule stacked by agent type}\label{fig:power_cs_electric}
    \end{subfigure}
    \begin{subfigure}{0.49\textwidth}
      \centering
      \includegraphics[width=1\textwidth]{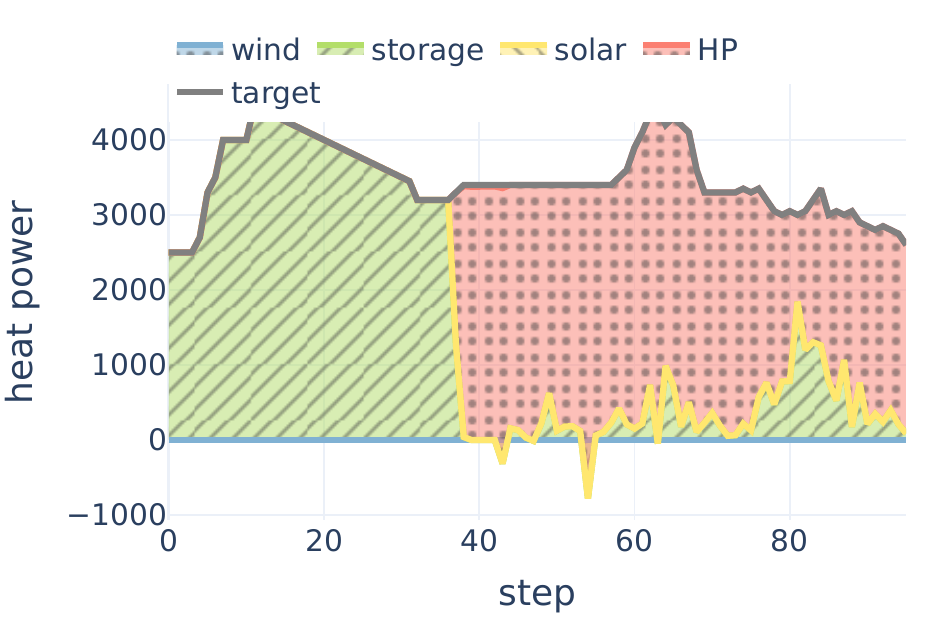}
      \subcaption{PES-H: Area plot of the heat cluster schedule stacked by agent type}\label{fig:heat_cs_electric}
    \end{subfigure}
    \caption{Examplary cluster schedules of the scenarios}\label{fig:cs_1}
\end{figure}
\begin{figure*}[htb]
    \centering
    \begin{subfigure}{0.28\textwidth}
        \centering
        \includegraphics[width=1\textwidth]{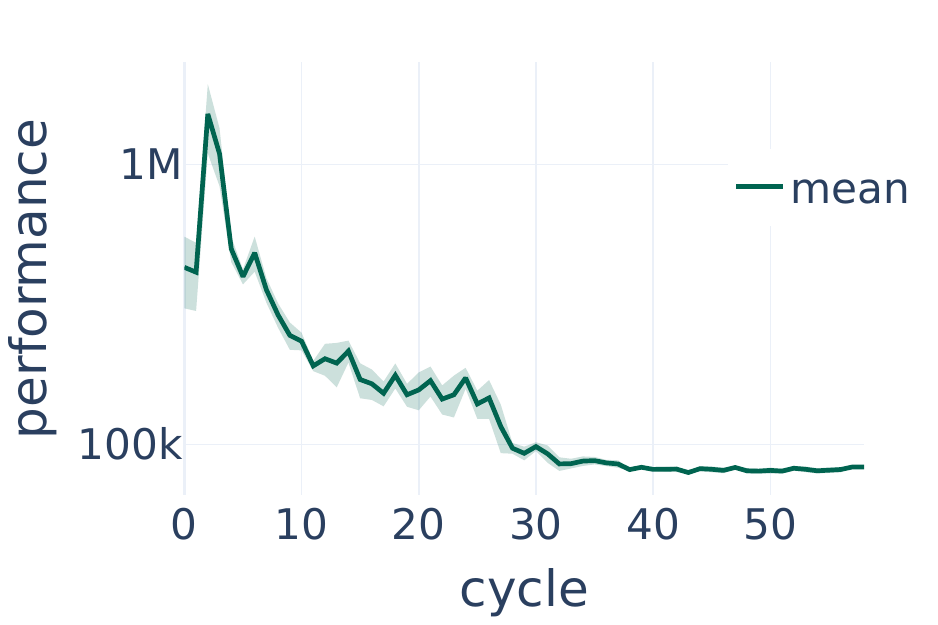}
        \subcaption{Convergence of GBS-L}\label{fig:convergence_CHP_es}
    \end{subfigure}
    \begin{subfigure}{0.28\textwidth}
        \centering
        \includegraphics[width=1\textwidth]{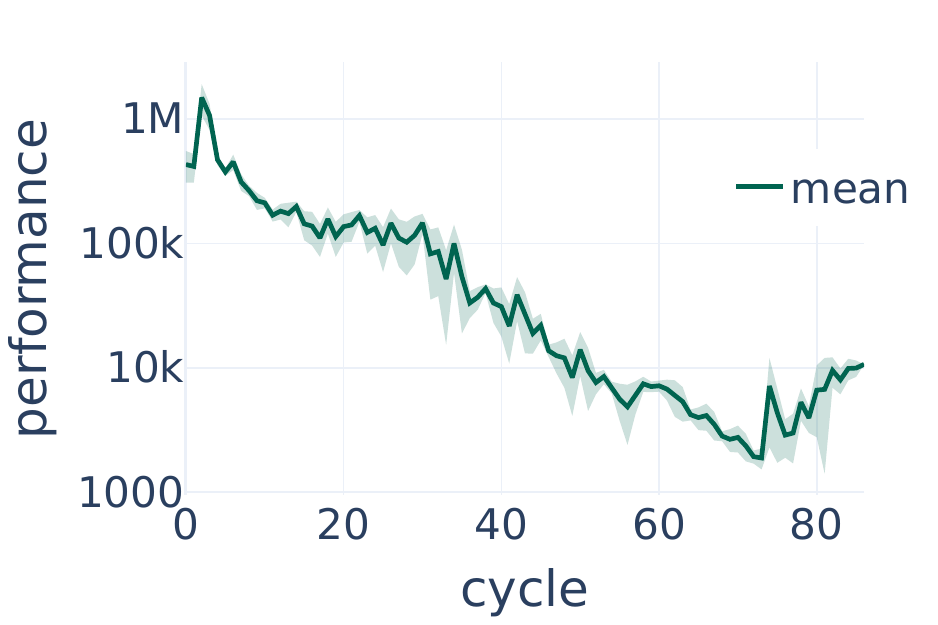}
        \subcaption{Convergence of GBS-H}\label{fig:convergence_CHP_es_hp}
    \end{subfigure}
    \begin{subfigure}{0.28\textwidth}
        \centering
        \includegraphics[width=1\textwidth]{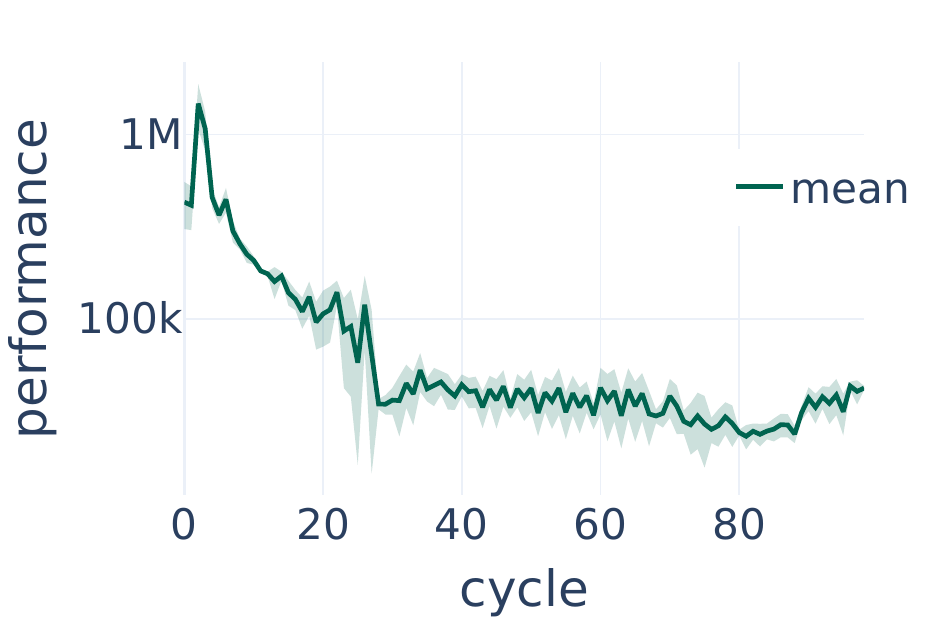}
        \subcaption{Convergence of the GBS-M}\label{fig:convergence_CHP_es_market}
    \end{subfigure}
    \begin{subfigure}{0.28\textwidth}
        \centering
        \includegraphics[width=1\textwidth]{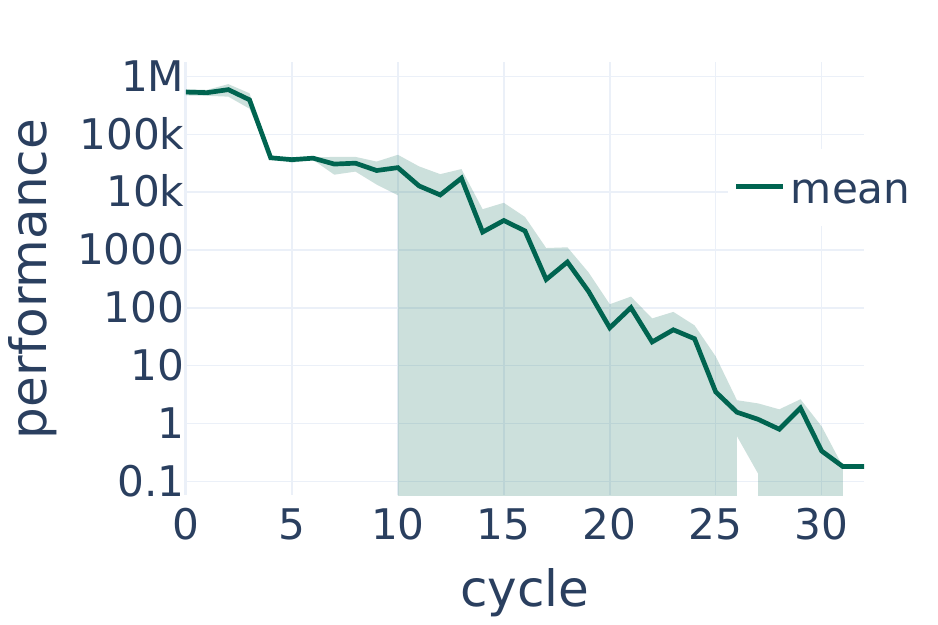}
        \subcaption{Convergence of the PES-L}\label{fig:convergence_hp_es}
    \end{subfigure}
    \begin{subfigure}{0.28\textwidth}
        \centering
        \includegraphics[width=1\textwidth]{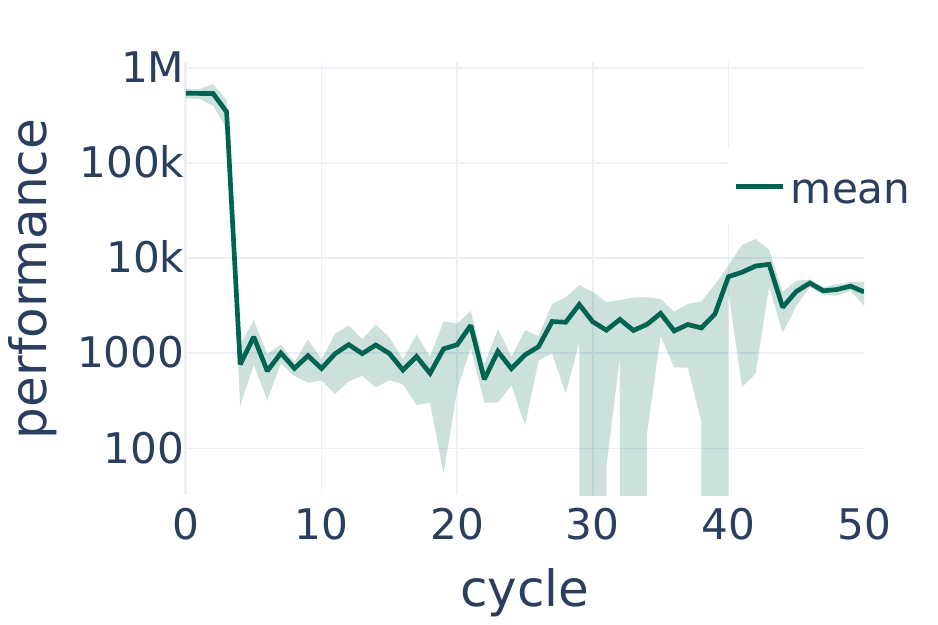}
        \subcaption{Convergence of the PES-H}\label{fig:convergence_hp_es_hp}
    \end{subfigure}
    \begin{subfigure}{0.28\textwidth}
        \centering
        \includegraphics[width=1\textwidth]{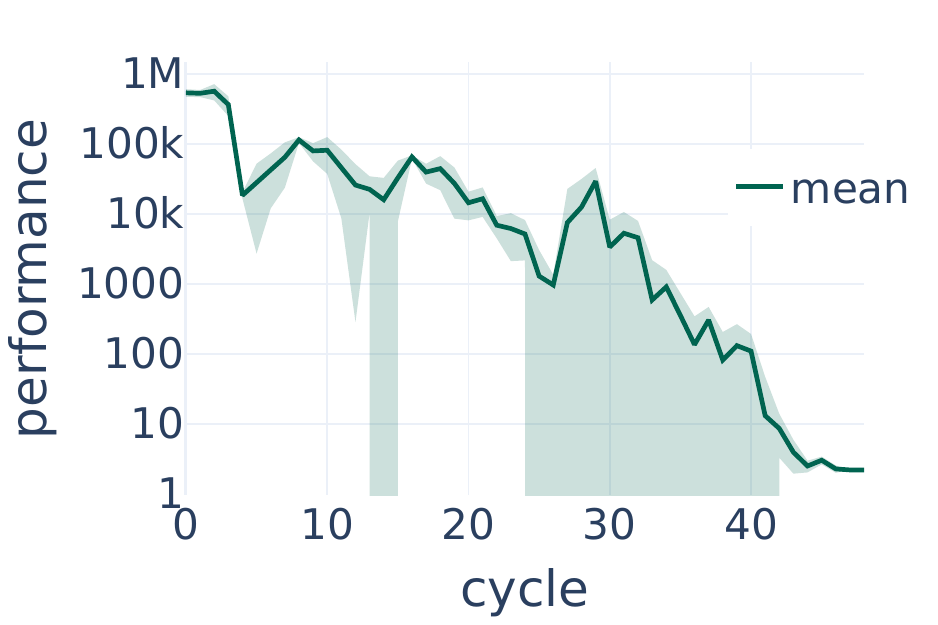}
        \subcaption{Convergence of the PES-M}\label{fig:convergence_hp_es_market}
    \end{subfigure}
    \caption{Convergence behavior}\label{fig:covergence}
\end{figure*}
\section{Results}
We present the results in four parts. First, we will introduce the results of the gas-based scenario without integrating storage. This shows the general problem, implied by having multiple sectors that need to be fulfilled, and demonstrates the necessity of storage inclusion for simultaneous optimization. After, the final results of the gas-based scenario are shown and discussed. Using the same target profiles, we will demonstrate how a pure-electrical solution for this problem could be solved using the appropriate renewable actors and adequate storage capacities. Further, we will show and discuss the results of the scenario variants regarding the different penalties and the utility function. Finally, we present the results on the execution time for each scenario.
\subsection{Gas-based system}
\begin{figure*}[htb]
    \centering
    \begin{subfigure}{0.48\textwidth}
        \centering
        \includegraphics[width=1\textwidth]{\detokenize{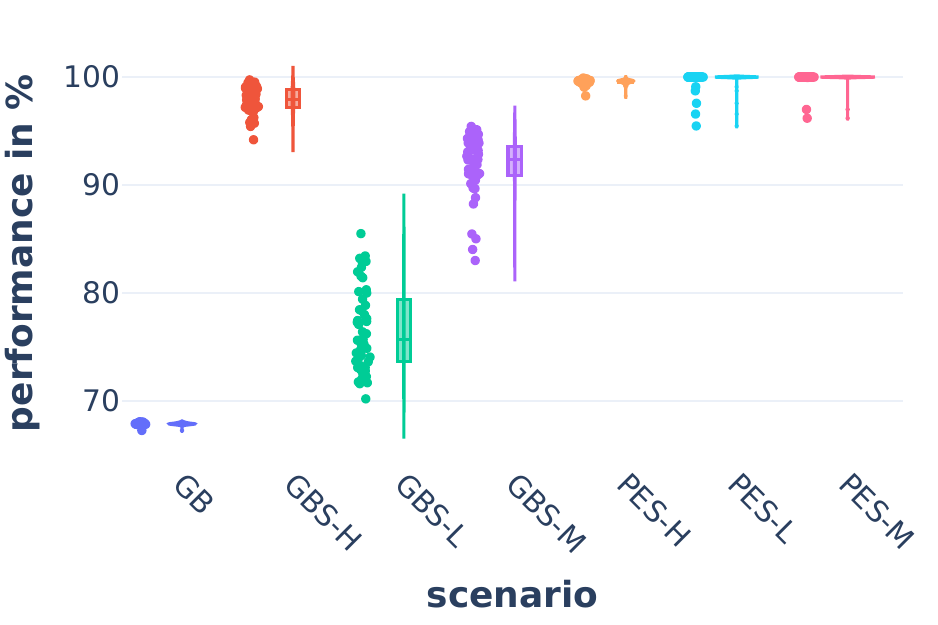}}
        \subcaption{Distribution of the collective performance in \% of 50 individual distributed optimizations for every scenario.}\label{fig:performance_distributions}
    \end{subfigure}
    \begin{subfigure}{0.48\textwidth}  
        \centering
        \includegraphics[width=1\textwidth]{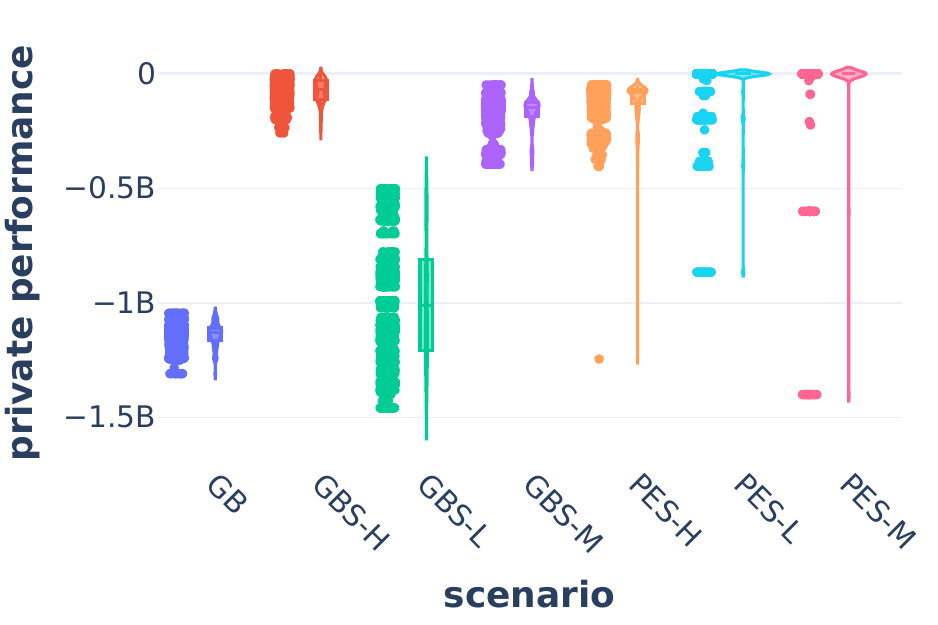}
        \subcaption{Distribution of the individual performances (no unit) of 50 individual distributed optimizations for every scenario and every agent.}\label{fig:performance_distributions_private}
    \end{subfigure}
    \begin{subfigure}{0.48\textwidth}  
        \centering
        \includegraphics[width=1\textwidth]{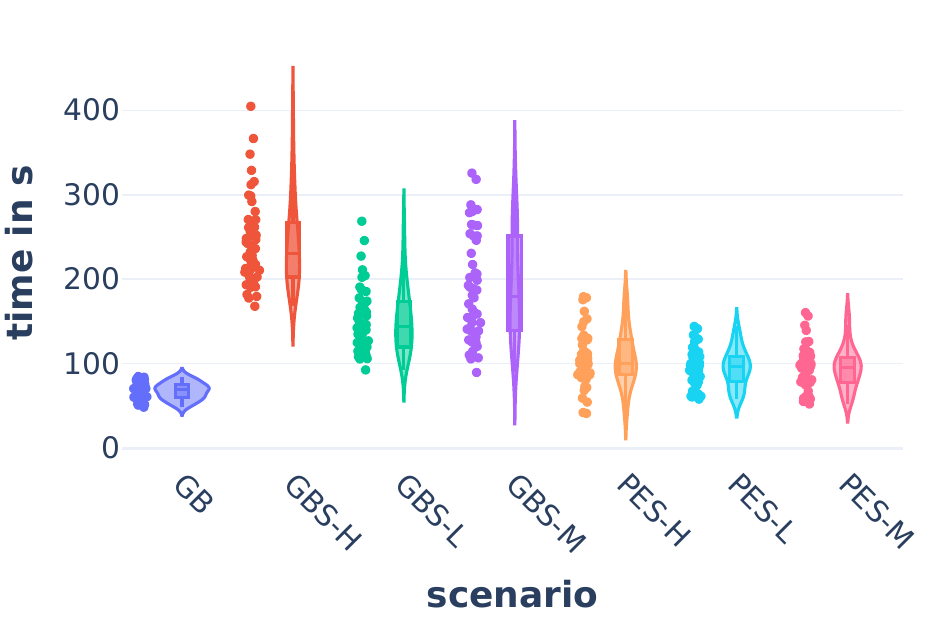}
        \subcaption{Execution time of every scenario}\label{fig:execution_time}
    \end{subfigure}
    \caption{Performance metrics and execution time}\label{fig:results_and_execution_time}
\end{figure*}

Before we look at the final results of the scenario described in the previous section, we shortly motivate the necessity of storage integration. We simulated the target profiles with the agents represented in Table \ref{tab:gas_scenario} excluding the storage agents. The results of this simulation are depicted in Figure \ref{fig:power_cs_industry} and \ref{fig:heat_cs_industry} for the stacked schedules. The results show that the optimization generated compromised \ac{CHP} schedules, which leads to over-fulfillment of the electrical target if the heat demand is relatively higher. In contrast, the thermal target is under-fulfilled at these times and vice versa (e.g., steps 10-30, 70-90, etc.). The implication is that it is impossible to operate such systems with these demand curves without curtailment of renewables (which economically are undesirable) or, second, storage capacities to balance these factors. This generally holds if energy should be balanced, especially locally (e.g., islanded grids, local energy communities). However, despite these problems, our approach leads to a fulfillment of slightly below 70\% (see Figure \ref{fig:performance_distributions}).

Integrating thermal and electrical storage agents makes the results look smoother and increases the fulfillment rate. The stacked schedules of an individual run are depicted in \ref{fig:power_cs_storage} and \ref{fig:heat_cs_storage}, the convergence behavior is shown in Figure \ref{fig:convergence_hp_es}, and the overall results as violin plots are depicted in Figure \ref{fig:performance_distributions}. These results show overall smooth cluster schedules with regard to the electrical and thermal target schedules. In the first section, both storages primarily provide the energy to account for the lack of solar energy, while the storages charge when the solar energy capability peaks. Further, the storage primarily balances the different peaks in the two energy sector targets. The convergence behavior is steady and does not experience a high variance over time. Further, the fulfillment rate is primarily near optimal; here, two outliers have a fulfillment rate below 95\%. The private objective also converges against zero as the penalties dominate the profit calculation.
\subsection{Pure-electrical system}
The overall results for the electrical system, which relies on \ac{HP}s, are depicted in Figure \ref{fig:performance_distributions} for the global performance, Figure \ref{fig:performance_distributions_private} for the private performance, and Figure \ref{fig:convergence_hp_es} for the convergence behavior. Figure \ref{fig:power_cs_electric} and Figure \ref{fig:heat_cs_electric} depict an individual stacked schedule.

It is visible that the fulfillment rate of the scenario is very high and very consistently high at around 99\%. The convergence behavior is jumpy and not monotonous, decreasing after the first five cycles. The variance is high in the lower performance area. However, the overall variance is low; the optimization approach can consistently find fitting schedules for all participants. When looking at the cluster schedule, it is noticeable that the storage contributes most of the energy when the solar and wind energy output is low and charges with the help of the solar power peak. The thermal storage primarily contributes initially, as the electrical supply capabilities are low due to the missing solar power output. Further, the electrical storage can only provide the initial electrical power and would not have been able to charge in the first ~30 steps.

It is further observable that the \ac{HP} agents recognize the relatively high difficulty of providing power before and after the sun's peak period. While it is not active initially, its output is regulated after the peak period, and the electrical storage has to step in for the \ac{SPP}.
\subsection{Influence of hyperparameters and utility coefficients}
Besides the approach as is, we also conducted some experiments on the base scenario variants to learn more about the influence of the penalty and the private utility function. Generally, it can be said that an even weighting of the economic approach with integrating the possibility for the agents to bid on the market (we used one day of March 2023 \cite{smard2023market}) leads to a globally suboptimal solution, while the agents, especially the \ac{CHP} agents, do not provide their power flexibility at the time of high market prices. Further, the thermal and electrical energy prices paid to the agent when contributing to the systems' objective can be used to weigh one sector over the other. This relation is nearly linear.

However, the results shown in Figure \ref{fig:performance_distributions} make clear that the inclusion of the market trading can act as a penalty replacement. The overall performance of the market scenarios is lower than the one with high penalties without the market but higher than that with standard penalties. We also observed that a combination (high penalties and the market) does not improve the results much further, so it is an act of balancing.
\subsection{Execution time}
One of the advantages of the proposed approach is the relatively low execution time, especially when compared against \textit{more complex} approaches like metaheuristics \cite{schrage2023multi, ikeda2015metaheuristic}. The execution time of all scenarios is shown in Figure \ref{fig:execution_time}. It is visible that the variance is low, but the average does differ a lot when looking at different scenarios. However, the absolute execution time is generally low enough considering the planning horizon of one day. Also, when evaluating the distributed nature of the simulation, the execution times are way below existing distributed multi-objective optimization, which also integrates storage \cite{schrage2023multi} ($>30$min, Pareto optimization). Still, it needs more time than central approaches \cite{ikeda2015metaheuristic} ($<< 1$min). Note that these measurements are only meant as a hint, as they depend on the specific hardware. The measurements have been done on the following CPU (using only a single core): \textit{AMD Genoa EPYC 9554 64Cores, 3.1 GHz, 360W}.
\section{Conclusion and Outlook}
This paper presented an approach to execute a distributed, multi-level, multi-objective optimization of plants across energy sectors. Our approach can simultaneously create a schedule in different scenarios to fulfill the thermal and electrical demand (target schedule). We showed a gas-based scenario with \ac{CHP}s with storage and without storage integration. Further, we simulated a pure-electrical scenario and demonstrated storage's role in these scenarios. At the same time, our simulation proves the feasibility of flexibility modeling and calculation of models like Amplify, even as part of a distributed optimization approach.

However, there are still several areas for improvement of the approach. First, it is relatively easy to choose appropriate hyperparameters due to the intuitive behavior of the approach. Still, it is necessary to set weighting coefficients to get valuable results. In the future, it might be better to lay out strict and clear rules (like strict prioritization) when formulating the optimization problem. Further, the local search has many advantages; it is fast and reliable. But it is also prone to finding plateaus when optimizing. To improve this, we would suggest trying non-population-based metaheuristic approaches and more local search derivates as it proved generally effective.

\backmatter

\section*{Glossary}

\begin{acronym}
\acro{CHP}[CHP]{combined heat and power}
\acro{ES}[ES]{energy storage}
\acro{COHDA}[COHDA]{Combinatorial Optimization Heuristic for Distributed Agents}
\acro{WPP}[WPP]{wind power plant}
\acro{SPP}[SPP]{solar power plant}
\acro{TES}[TES]{thermal energy storage}
\acro{VPP}[VPP]{virtual power plants}
\acro{EH}[EH]{energy hub}
\acro{ADMM}[ADMM]{alternating direction method of multipliers}
\acro{DGD}[DGD]{distributed gradient descent}
\acro{RER}[RER]{renewable energy resource}
\acro{DER}[DER]{distributed energy resource}
\acro{HP}[HP]{heat pumps}
\end{acronym}

\section*{Declarations}

\subsection*{Funding} 
This work has been funded by the Deutsche Forschungsgemeinschaft (DFG, German Research Foundation) – 359941476.

The simulations were performed at the HPC Cluster ROSA, located at the University of Oldenburg (Germany) and funded by the DFG through its Major Research Instrumentation Programme (INST 184/225-1 FUGG) and the Ministry of Science and Culture (MWK) of the Lower Saxony State.

\subsection*{Competing interests}
The authors declare that they have no competing interests.

\subsection*{Availability of data and materials}
The code and the data used as input for the simulations are available at \url{https://github.com/Digitalized-Energy-Systems/heat-electricity-storage-cohda}.

\subsection*{Author contribution}
RS was responsible for creating the initial paper draft, conceptual work, parts of the implementation, the execution of the simulation, and supervision. JR developed the initial draft of the concept, the initial implementation, and the simulations. AN was responsible for quality assurance and for supervising the work as a whole.

\section*{Disclaimer}
The manuscript was submitted to Springer Open Energy Informatics and rejected. We do not have the resources to rework the manuscript shortly, so we uploaded it after a minor revision. 

\begin{appendices}




\end{appendices}


\bibliography{main}

\end{document}